\documentclass[a4paper]{article}

\usepackage{icrc2013}

\title{A Very Local Interstellar Spectrum for Galactic Electrons}

\shorttitle{Local Interstellar Spectrum for Electrons}

\authors{M.S. Potgieter$^{1}$, E.E. Vos$^{1}$, R.R. Nndanganeni$^{1}$, M. Boezio$^{2}$, R. Munini $^{2}$ 
}

\afiliations{
$^1$Centre for Space Research, North-West University, Potchefstroom, South Africa\\ 
$^2$INFN, Sezione di Trieste, Trieste, Italy
}

\email{Marius.Potgieter@nwu.ac.za}

\abstract{We present a new local interstellar spectrum for cosmic ray electrons over an energy range from 1 MeV to 70 GeV. Below ($0.8\pm0.2$) GeV it has a power law form, $E^{-(1.55\pm0.05)}$, with $E$ kinetic energy, which is consistent with previous studies. 
This is derived from comparing Voyager 1 electron data observed during 2010 with a comprehensive modulation model. 
However, reproducing the PAMELA electron spectrum observed at Earth during late 2006 and to address an unexpected increase in the electron spectrum between $\sim$2 GeV 
and up to $\sim$20 GeV, a spectral index of $-(3.3\pm0.1)$ instead of the reported $-(3.18\pm0.05)$ was found for this energy range. 
This feature cannot be caused by solar modulation or any other process inside the heliosphere.}


\begin{document}
\maketitle

\section{Introduction}

Modelling of the solar modulation of cosmic rays (CRs) requires that an input spectrum must be specified at an assumed modulation boundary. This spectrum, called a local interstellar spectrum (LIS), is then modulated throughout the heliosphere as a function of position, energy and time. 
Computed modulated spectra are usually compared to observations close to or at Earth and along trajectories of spacecraft such as Voyager 1 and 2, Ulysses and others to study the solar modulation process in fine detail.
	
This process is determined by a combination of modulation parameters including the solar wind speed and its spatial dependence, the heliospheric magnetic field with its embedded turbulence and global geometry, gradient, curvature and current sheet drifts, adiabatic energy changes and of course the geometry of the heliosphere and where the modulation boundary is located [1,2]. 
Establishing the three major diffusion coefficients and particle drifts, as a function of time, space and energy, is a work in progress. Since these parameters are not uniquely known, the LIS cannot be determined by using only observations at or close to Earth (despite the user friendliness of the force-field modulation approach as applied e.g. by [13]).

Most helpful in this context is that CR observations are now becoming available from Voyager 1 and 2, both positioned deep inside the inner heliosheath, and with Voyager 1 probably already in the heliopause region [3,4]. 
When these CR spectra observed in the outer heliosphere, and at Earth [5,6], are compared to computed spectra over the full energy range relevant to solar modulation, (from 1 MeV to 50 GeV), the modulation parameters can be better determined so that trustworthy conclusions about the input spectrum can be made.
	
The required input spectrum is in fact a heliopause spectrum (HP), not a galactic spectrum. In this context it should be mentioned that computed galactic CR spectra (GS) do not usually contain the contributions of any specific (local) sources within parsecs from the heliosphere 
so that an interstellar spectrum may be different from an average GS which may be different from a LIS (thousands of AU away from the Sun), which might be different from a very LIS or what can be called a heliopause spectrum, right at the edge of the heliosphere, say within  $\sim$200 AU away from the Sun. If known, the latter would be the ideal spectrum to use as an input spectrum for solar modulation models.

Attempting to make progress in this regard, the solutions of a comprehensive numerical model (not the force-field modulation approach - see [13]) for the modulation of galactic electrons in the heliosphere are presented in comparison with Voyager 1 and PAMELA observations [5,6[. This is done to obtain a computed electron spectrum at the heliopause that can be considered the lowest possible very LIS.
 
This report follows up on the work by Potgieter and Nndanganeni [1] and Nkosi et al. [2] on the solar modulation of electrons in the heliosphere, especially in the outer heliosphere. It contains observational data from Voyager 1 [3,4] and the PAMELA mission [5,6], with reference to the detailed spectra from PAMELA as reported by Di Felice [7], Munini [8] and Vos [9]. The finalized electron spectra from PAMELA for 2006 to 2009 should become available soon. See also supporting and complimentary reports on observations and modelling at this ICRC conference.  

\begin{figure*}[!t]
\centering
\includegraphics[width=0.75\textwidth]{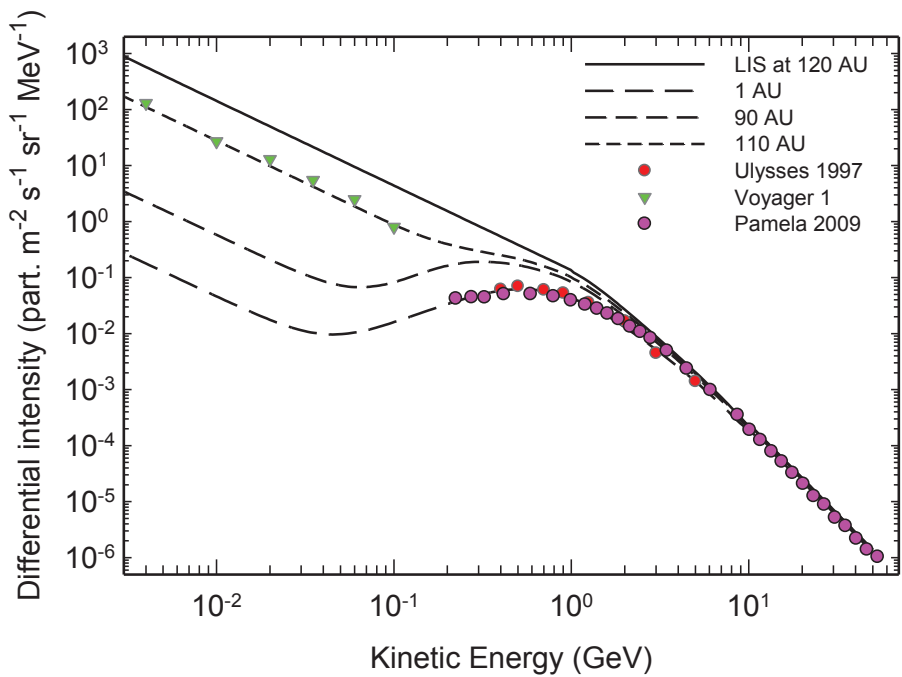}
			\caption{Computed modulated electron spectra at Earth (with polar angle of $\theta = 90^{\circ}$) at 90 AU and 110 AU 
(with $\theta = 60^{\circ}$ corresponding to the Voyager 1 trajectory), showing how they are modulated from the inner to the outer heliosphere. 
These spectra are completely compatible to observations from Voyager 1 during 2010, PAMELA during 2009 [5,6,7,8,9], and from Ulysses as observed in 1997 [12]. 
This very LIS is specified at the heliopause positioned at 120 AU. See also [1,14].}
\label{fig:Figure1}
\end{figure*}

\section{Numerical Model}

A full three-dimensional (3D) numerical model, with three spatial dimensions and an energy dependence (four computational dimensions), was used to compute electron spectra at selected positions in the heliosphere, including the inner heliosheath. 
The details of this model were published by [1,9] and are not repeated here. For an equivalent model applied to proton modulation, 
see [10]. The model is based on the numerical solution of a full 3D transport equation [11] for solar modulation including all four major modulation processes with a full diffusion tensor [10]. 
The focus is on solar minimum modulation. We assume that solar modulation becomes negligible with $E > 50$ GeV.
For a calculation at what rigidity the solar modulation of CRs begins, see Potgieter and Strauss (this ICRC volume).

The heliopause is varied between 120 to 122 AU in the model with the TS at 94 AU, which gives a 26 to 28 AU wide heliosheath in the direction in which Voyager 1 has been moving. This width is a determining factor in the modulation of galactic electrons between 1 MeV and 100 MeV [1,2,9].

\section{Results and Discussion}

The numerical model and the available Voyager 1 observations of galactic electrons between 6 MeV to $\sim$120 MeV during 2010 is used to determine a heliopause spectrum which can be considered a very local interstellar spectrum below 100 MeV. 
The electron spectrum observed in 2010 is an optimal choice from the available Voyager 1 data because earlier observed spectra, closer to the TS, are much lower and may be subjected to short-term changes as the region closer to the TS is more turbulent as it shifts position with changing solar activity.

The input spectrum, as well as the modulation parameters, were adjusted to reproduce the Voyager spectrum observed during 2010 in terms of differential flux values between 1 MeV and 200 MeV. 
The spectral slope was then adjusted with a different power law index above $\sim$1 GeV to reproduce the PAMELA data at Earth. These spectra unfortunately extend down to only  $\sim$200 MeV whereas the Voyager 1 measurements extend only up to $\sim$120 MeV. All The modulation parameters as required to reproduce the Voyager 1 and PAMELA observations were given and motivated by [1,9].

First, the results and very LIS are shown in figure 1. It is based on a heliosphere with a TS at 94 AU and a HP at 120 AU. The solid line is the LIS which, when modulated as described above, reproduces the Voyager 1 observations very well and simultaneously also the PAMELA electron spectrum at Earth for the end of 2009 and the Ulysses spectrum for 1997 [12,15]. 
Although Ulysses was not at Earth, the spectra are quite similar because very small latitudinal gradients had been reported [12]. 
To reproduce the spectral shape of the Voyager 1 spectrum, the LIS below $\sim$1.0 GeV must have a power law form with
$E^{-(1.55\pm0.05)}$, where $E$ is kinetic energy. To reproduce the PAMELA electron spectra observed between the end of 2006 and the end of 2009, the LIS must have a power law form with $E^{-(3.13\pm 0.05)}$, above $\sim$5 GeV. This LIS thus consists of two power laws with a 'break' occurring between
$\sim$800 MeV and $\sim$2 GeV. Note that [1,14] reported a power law with a spectral index of $-(3.15\pm 0.05)$,

\begin{figure*}[t]
\centering
\includegraphics[width=0.75\textwidth]{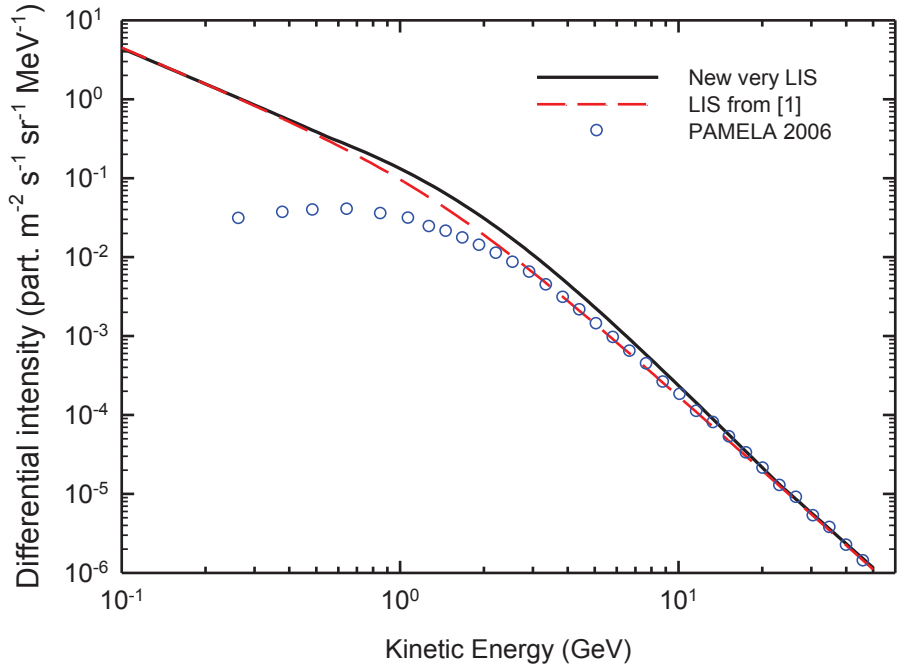}
			\caption{A comparison between the electron LIS reported here (solid black line) and a LIS (dash line in red) with a 
spectral index of $−(3.13\pm0.05)$ above 2 GeV and normalized to the PAMELA data above 30 GeV. The preliminary PAMELA electron intensities from November 2006 are presented by the symbols.  At lower energies the LIS from [1] is used. Note that data between $\sim$2 GeV and $\sim$10 GeV, perhaps even 20 GeV, requires a different spectral slope than above 30 GeV.}
\label{fig:Figure2}
\end{figure*}

Concerning this break in the spectral shape of the presented HP spectrum, Strong et al. [13] re-analysed synchrotron radiation data using various radio surveys to constrain the low-energy electron LIS in combination with data from Fermi-LAT and other experiments. 
They concluded that the electron LIS exhibits a spectral break below a few GeV in accord with what is presented here. 
In addition they stated that the LIS below a few GeV has to be lower than what standard galactic propagation models predict, again consistent to what is shown here, and speculated about the causes of this result. 
It should be noted that the relation between synchrotron data and electrons is complicated by the presence of secondary electrons and positrons in the Galaxy so that it is not a straightforward matter to determine the spectral slope below a few GeV with galactic propagation modelling.

Further study of the preliminary electron spectra from PAMELA for the period 2006 to 2009 as reported by [9] identified a new feature in the electron spectrum which is not evident from figure 1. 
This new feature is shown in figure 2 where a comparison is made between the PAMELA spectrum for November 2006 and a LIS with a spectral index of -3.15 above 2 GeV and normalized to the PAMELA data above 30 GeV as in figure 1. 

According to Adriani et al. [6], who conducted a study of electron measurements made by PAMELA from July 2006 to January 2010, between $1\textnormal{ GeV}$ and $625\textnormal{ GeV}$, a spectral index of $-(3.18\pm 0.05)$ is required for electrons above 30 GeV. 

Shown in figure 2 is that the LIS from figure 1 is clearly below the observed differential flux values observed at Earth between $\sim$2 GeV and $\sim$10 GeV.  Of course, a LIS that is lower than the flux at Earth is unacceptable from a modulation point of view. 
At these energies, the modulation is already large enough so that the observed values should always be below the assumed LIS. 
Clearly,  the $E^{-3.15}$ dependence at higher energies is obscured by an unusual region of enhanced intensities observed between 
$\sim$2 GeV and up to $\sim$20.0 GeV. 
This means the LIS must be adjusted in this energy range to have a different spectral index than the reported -3.15. The $E^{-3.15}$ dependence is only visible above $\sim$20 GeV. The LIS shown as the solid black line takes this feature of the observed data into account. 

The new LIS, taking this effect into account, is given by: 

\begin{equation}
 j_{LIS} = \frac{0.132}{\beta^2}\left[\frac{\left(\frac{E}{E_L}\right)^{1.6}+\left(\frac{E_{k_1}}{E_L}\right)^{1.6}}
{1+\left(\frac{E_{k_1}}{E_L}\right)^{1.6}}\right]^{\frac{b_1-a_1}{1.6}}\left(\frac{E}{E_L}\right)^{a_1}
\end{equation}

\textnormal{for $0.55\textnormal{ GeV}\leq{E}\leq22.70\textnormal{ GeV}$}, and 

\begin{equation}
 j_{LIS} = \frac{0.118}{\beta^2}\left[\frac{\left(\frac{E}{E_L}\right)^{3.0}+\left(\frac{E_{k_2}}{E_L}\right)^{3.0}}
{1+\left(\frac{E_{k_2}}{E_L}\right)^{3.0}}\right]^{\frac{b_2-a_2}{3.0}}\left(\frac{E}{E_L}\right)^{a_2} 
\end{equation}

\textnormal{elsewhere,} with $E_L$=1 GeV taking care of the units, and where $E_{k_1}=1.7\textnormal{ GeV}$, $E_{k_2}=1.4\textnormal{ GeV}$, 
\mbox{$a_1=-1.00$}, $b_1=-3.54$, $a_2=-1.50$, and $b_2=-3.13$ are parameters that determine the shape of the LIS.

The question arises of what can cause such an increase in the electron spectrum below about 10 GeV while a steady decrease is expected because of increasing solar modulation. 
Since no process in the heliosphere is able to accelerate CR particles up to such high energies, it is assumed that the appearance of such an unusual “bump” in the spectrum could most likely be ascribed to a process or more likely the presence of a local source effect in the interstellar medium. 
Further investigation from an astrophysics point of view seems required.

Modulation beyond the HP has become a very relevant topic since both Voyager spacecraft are about to explore the 
outer heliosheath (beyond the HP) and therefore may actually measure a pristine LIS sooner than what we anticipate. 
Recently, Strauss et al. [16] computed that the differential intensity of 100 MeV protons may decrease by $\sim 25\%$
from where the heliosphere is turbulently disturbed (inwards from the heliospheric bow wave) up to the HP. However, because the diffusion coefficients 
of low energy electrons are independent of energy, making them significantly larger than for protons of the same rigidity, this percentage should be 
much less for 100 MeV electrons. It could be that we underestimate this effect and that the true electron LIS below $\sim$200 MeV is higher than what we present here. 
It appears from counting rates (not differential flux) reported by [4] that Voyager 1 observed a significant increase at what appears to be the 
heliopause. Future Voyager 1 observations should enlighten us in this respect as it moves outwards at 3 AU per year. It should also be kept in mind 
that the Voyager 1 and 2 detectors cannot distinguish between electrons and positrons whereas the PAMELA detector can. (See also Potgieter et al., this conference). 

\section{Conclusions}

We present a new LIS for cosmic ray electrons over an energy range from 1 MeV to 50 GeV. 
Below $\sim$1.0 GeV this LIS has a power law form with $E^{-(1.55\pm0.05)}$ consistent with the LIS shown in figure 1 and 
derived by [1,14, also this conference] from comparing computations with the Voyager 1 electron spectrum from 2010. 
During this year, Voyager 1 had moved from 112 AU to 115 AU.

However, in order to reproduce the PAMELA electron spectrum for the end of 2006, the LIS had to be adjusted to address an unexpected ``bulge'' 
in the electron spectrum between $\sim$2 GeV and up to $\sim$20 GeV. This produces a spectral index of $-(3.3\pm 0.1)$ instead of 
$-(3.13\pm 0.05)$ in the mentioned energy range. 

This feature cannot be caused by solar modulation or any other process inside the heliosphere, such as the reacceleration of galactic electrons at the solar wind termination shock.
  
\vspace*{0.5cm}
\footnotesize{{\bf Acknowledgment:}{~The authors thank Bill Webber for providing them with the electron data from Voyager 1. 
The partial financial support of the South African National Research Foundation (NRF), 
the SA High Performance Computing Centre (CHPC) and the SA Space Agency's (SANSA) Space Science Division is acknowledged.}

\end{document}